\documentstyle[preprint,tighten,epsfig,pra,aps]{revtex}
\begin{document} \draft

\title{\LARGE \bf Quantum Chaos at Finite Temperature - a New Approach via the Quantum Action }

\author{ L.A. Caron, H. Jirari, H. Kr\"oger, G. Melkonyan }
\address{D\'epartement de Physique, Universit\'e Laval, Qu\'ebec, Qu\'ebec G1K 7P4, Canada}

\author{ X.Q. Luo }
\address{Department of Physics, Zhongshan University, Guangzhou 510275, China}

\author{ K.J.M. Moriarty }
\address{Department of Mathematics, Statistics and Computer Science, 
Dalhousie University, Halifax, Nova Scotia B3H 3J5, Canada}

\footnotetext{talk given by H. Kr\"{o}ger, email: hkroger@phy.ulaval.ca}

\maketitle

\begin{abstract}
We address the problem of quantum chaos: Is there a rigorous, physically meaningful definition of chaos in quantum physics? Can the tools of classical chaos theory, like Lyapunov exponents, Poincar\'e sections etc. be carried over to quantum systems? Can quantitative predictions be made?
We show that the recently proposed quantum action is well suited to answer those questions. As an example we study chaotic behavior of the 2-D 
anharmonic oscillator and compare classical with quantum chaos. Moreover, we study quantum chaos as function of temperature (the classical system can be considered as the limit where temperature goes to infinity).
\end{abstract}

\section{Introduction}
Heisenberg's uncertainty principle, quantum tunneling, Schr\"odinger's cat paradox, entangled states, Einstein-Rosen-Podolski paradox, quantum cryptology, quantum computing are presently hot topics in quantum physics, 
which are intrinsically of quantum nature, i.e. have no analogon in classical physics.
On the other hand, in modern quantum physics there are concepts 
which have their origin in classical physics.
Examples are instantons and chaos.
Chaotic phenomena in quantum systems have been identified 
in few-body systems of atomic or molecular systems.
For an overview see Ref. \cite{Blumel}.
For example, the hydrogen atom in a strong magnetic field shows strong irregularities in its spectrum \cite{Friedrich}.
Other systems, which have attracted much attention are traps.
An example is the Paul-trap, where trapping of two ions can be described by a 
simple Hamiltonian \cite{Paultrap}.
Another example of chaos in quantum physics is a billiard, where a single particle interacts with the walls (in a model). Experimentally, the wall can be realized by a chain of atoms forming a corral \cite{Stockmann}, or even by laser light. Quite recently the motion of ultra-cold atoms in a billard formed by laser beams has been experimentally realized by  Milner et al.\cite{Milner:01} and Friedman et al.\cite{Friedman:01}.    
This is an example of chaos in a many-body system.
The trace of quantum chaos has been observed long ago in irregular patterns in the wave functions of the quantum mechanical model of the stadium billard \cite{McDonald}.
Footprints of quantum chaos have been identified also in spin systems,
e.g. in the spectrum of lattice spin systems \cite{Nakamura:85}.
For an overview see Ref. \cite{Nakamura}. Also chaotic behavior of spin glass clusters has been investigated by Georgeot and Shepelyansky \cite{Georgeot}.

On the theoretical side, in order to describe chaos in quantum systems quantitatively, the following developments were successful.
One route has been to characterize the spectral density of quantum system with chaotic classical counterpart by Poisson versus Wigner distributions. There is a conjecture by Bohigas et al. \cite{Bohigas}, which says that the signature of a classical chaotic system is a spectral density following a Wigner distribution. 
Another successful approach was Gutzwiller's trace formula \cite{Gutzwiller}, which establishes a relation between Q.M. transition amplitudes and classical periodic orbits. In particular, it has been tested and found to work in the semiclassical regime of atoms (highly excited Rydberg states) \cite{Wintgen:88}.
However, recently an experimental test of a generalized trace formula applied to a microwave billiard realized by superconducting 2-dim microwave resonators has shown some deviation between theory and experiment \cite{Dembrowski:01}.

Unlike in classical chaos, where local information on the trajectories in phase space allows to characterize chaotic behavior, such information is  neither obtained from the spectral density nor from the trace formula.
According to common folklore, such information is principially not available in Q.M. The standard argument for this is Heisenberg's uncertainty relation, absent in classical physics.
Nevertheless, attempts have been made to describe the dynamics of quantum systems in as close as possible analogy to classical physics.
Such a route has been explored by Cametti et al. \cite{Cametti:99}, using the concept of the effective action.
In Q.M. de Broglie has pointed out the duality between the particle and wave interpretation. But there is a more modern view point: Renormalisation and the effective action. It means that the action of a quantum system can be written
like the action of the classical system, however, with modified parameters (mass, potential parameters).  
The concept of renormalisation and the effective 
action has been useful to describe an equivalent low-energy theory, starting from a high-energy theory. For example, the construction of a low-energy theory in nuclear physics has been discussed by Lepage \cite{Lepage}.
The effective action $\Gamma$
has been introduced in quantum field theory \cite{Jona,Coleman}, giving an expectation value $<\phi>$ which minimizes the potential energy, giving the ground state energy. 
The concept of effective action has been generalized also to include finite temperature effects \cite{Dolan}.

Let us consider the effective action in quantum mechanics, computed using perturbation theory (loop expansion) by  
Cametti et al. \cite{Cametti:99}. They consider the Lagrangian
\begin{eqnarray}
L(q,\dot{q},t) &=& \frac{m}{2} \dot{q}^{2} - V(q), ~~~ 
V(q) = \frac{m}{2} \omega^{2} q^{2} + U(q) ~ .
\end{eqnarray}
Here $U(q)$ is a local potential, e.g., the quartic potential $U(q) \sim q^{4}$. Then effective action obtained has the following form,
\begin{eqnarray}
\Gamma[q] &=& \int dt \left( - V^{eff}(q(t)) \right.
\nonumber \\
&+& \left. \frac{Z(q(t))}{2} \dot{q}^{2}(t)
+ A(q(t)) \dot{q}^{4}(t) + B(q(t)) (d^{2}q/dt^{2})^{2}(t) + \cdots \right)
\nonumber \\
V^{eff} &=& \frac{1}{2}m \omega^{2} q^{2} +U(q) + \hbar V^{eff}_{1}(q) + O(\hbar^{2})
\nonumber \\
Z(q) &=& m + \hbar Z_{1}(q) + O(\hbar^{2})
\nonumber \\
A(q) &=& \hbar A_{1}(q) + O(\hbar^{2})
\nonumber \\
B(q) &=& \hbar B_{1}(q) + O(\hbar^{2}) ~ .
\end{eqnarray}
One observes that the effective potential $V^{eff}$ as well as the mass renormalisation $Z$ is given by a series of increasing order loop corrections (proportional to powers of $\hbar$). The most important property is the 
occurrence of higher order time derivative terms in the kinetic term, corresponding to an asymptotic infinite series of increasing order. If one wants to interpret $\Gamma$ as effective action, and in particular use it to compute trajectories taking into account quantum corrections,
one is faced with a conceptual problem: The higher time derivatives require
more intial/boundary conditions than the classical action. Such initial data are not available. A possible way out is an approximation of the effective action to low order in $\hbar$. This has been done in Ref. \cite{Cametti:99} and quantum chaos of the 2-D anharmonic oscillator has been studied. 
In the following we will present an alternative way, via the quantum action.

\section{Quantum action}
In Ref.\cite{Jirari:a} the concept of the quantum action has been introduced.
The idea is that the a single quantum action gives a global fit of Q.M. transition amplitudes for a fixed transition time $T = t_{fi} - t_{in}$ and all possible combinations of initial and final boundary points $x_{in}$, $x_{fi}$.
Let us consider the Q.M. transition amplitude given by the path integral,
\begin{equation}
\label{QMTransAmpl}
G(x_{fi},T;x_{in},0) =
\left. \int [dx] \exp[ \frac{i}{\hbar} 
S[x] ] \right|_{x_{in},0}^{x_{fi},T} ~ ,
\end{equation}
where 
\begin{equation}
S = \int dt  \frac{m}{2} \dot{x}^{2} - V(x) 
\end{equation}
denotes the classical action. The quantum action 
is defined as follows. For the given classical action 
there is a quantum action, 
\begin{equation}
\tilde{S} = \int dt \frac{\tilde{m}}{2} \dot{x}^{2} - \tilde{V}(x) ~ ,
\end{equation}
which allows to express the transition amplitude by
\begin{equation}
\label{DefRenormAction}
G(x_{fi},T; x_{in},0) = \tilde{Z} 
\exp [ \frac{i}{\hbar} \left. \tilde{S}[\tilde{x}_{cl}] 
\right|_{x_{in},0}^{x_{fi},T} ] ~ .
\end{equation}
Here $\tilde{x}_ {cl}$ denotes the classical path, such that the action $\tilde{S}[\tilde{x}_{cl}]$ 
is minimal (we exclude the occurrence of conjugate points or caustics). 
$\tilde{Z}$ denotes the normalisation factor corresponding to $\tilde{S}$. 
Eq.(\ref{DefRenormAction}) is valid with 
the {\em same} action $\tilde{S}$ for all sets of 
boundary positions $x_{fi}$, $x_{in}$ for a given time interval $T$.  
Any dependence on $x_{fi}, x_{in}$ enters via the trajectory $\tilde{x}_ {cl}$. $\tilde{Z}$ is independent of $x_{fi}, x_{in}$, but depends on $T$.
The parameters of the quantum action depend on the transition time $T$.  
If we wish, we can choose to go over from real to imaginary time. An imaginary, finite transition time can be identified with an inverse temperature $\beta$ via
$\beta = T/\hbar$ and $\beta$ being related to the temperature $\tau$ by $\beta = 1/(k_{B} \tau)$. This leads to the interpretation that the parameters of the quantum action are a function of temperature.

The quantum action has been explored numerically in 
Ref.\cite{Jirari:a,Jirari:b,WCNA,Jirari:c,Caron}. 
In Ref.\cite{Jirari:b} the quantum action has been computed for a 1-D double well potential.
The quantum action allows to incorporate Q.M. fluctuations, which manifest themselves by action parameters (mass $\tilde{m}$, parameters of potential $\tilde{V}$) different from its classical counter part. 
Such a phenomenon is well known from relativistic quantum field theory, 
where the action parameters of 
the non-interacting theory are called bare parameters and those of the interacting field theory are called renormalized parameters.
The situation distinguishing classical physics from quantum physics is quite analogous: The noninteracting field theory has its analogue in classical physics. The role of bare parameters is played by classical action parameters. 
The interacting field theory has its analogue in quantum physics. The role of renormalized parameters is played by the parameters of the quantum action.  
Thus we see that the quantum action has a very physical interpretation as a renormalized theory, where the dynamical effect of quantum fluctuations is 
condensed into the parameters of the quantum action.

The double well potential is physically interesting because it allows to address the phenomenon of tunneling, and allows to study quantum instantons and compare those with classical instantons (see Coleman \cite{Coleman:85}).
Classical instantons are solutions of classical equations of motion in imaginary time, which for a classical potential with degenerate minima go from one 
extremum (maximum in imaginary time) to the other.  
What do we mean by a quantum instanton?
As an example of a system with a double-well potential, we have considered in 1-D the classical action with $m=1$, and 
$V(x)=v_{0} + v_{2} x^{2} + v_{4} x^{4}$ with $v_{0}=v_{4}=\frac{1}{2}$, $v_{2}=-1$. The parameters of the quantum action have been determined 
numerically in such a way, that the quantum action fits globally Q.M. transition amplitudes for a number of combinations of boundary points $x_{in}$, $x_{fi}$. 
The numerical results for the parameters as function of $T$ are shown in Fig.[1].
One observes that in the limit of large time $T$, the action parameters converge to some asymptotic limit. Because $T \to \infty$ corresponds to temperature $\tau \to 0$, this regime corresponds to the low temperature regime. For large imaginary time Q.M. transition amplitudes are dominated by the ground state wave function and ground state energy (Feynman-Kac formula).
Thus in the zero temperature limit, the parameters of the quantum action describe the quantum physics of the ground state. Analytical behavior of the quantum action in this regime has been discussed in Ref.\cite{Jirari:c}.
Now let us come back to the quantum instantons. As can be seen from Fig.[1b], the quadratic term $\tilde{v}_{2}$ of the quantum potential is negative for all $T$. Thus the quantum potential $\tilde{V}(x)$ has a double well shape and gives rise to instanton solutions. Those instanton solutions depend not only 
\begin{figure}
\begin{center}
\epsfig{figure=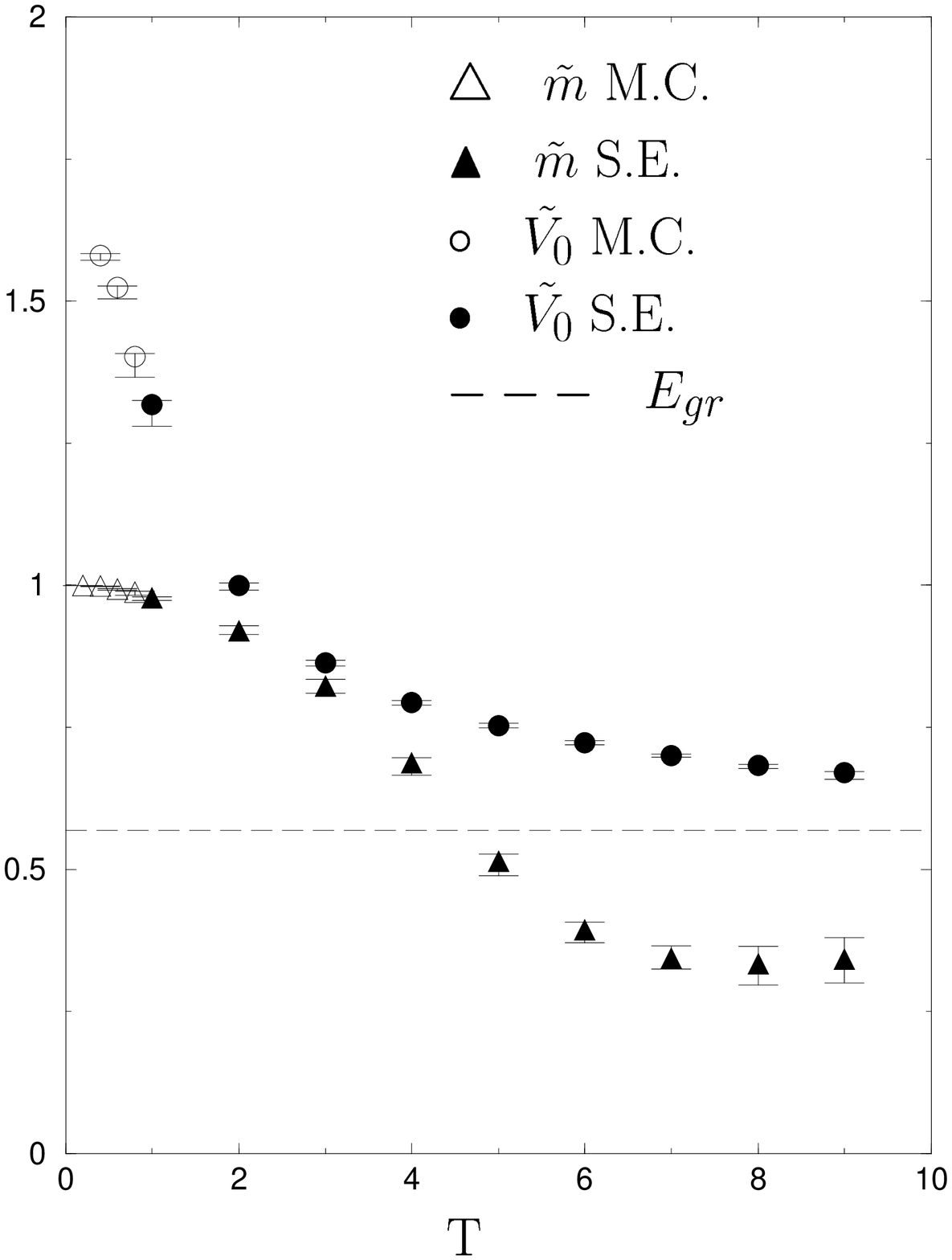,height=9cm,width=.9\linewidth,angle=0}
\end{center}
\noindent 
Fig.[1a] Parameters $\tilde{m}$ and $\tilde{v_0}$ of quantum action versus time $T$.
\end{figure}
\begin{figure}
\begin{center}
\epsfig{figure=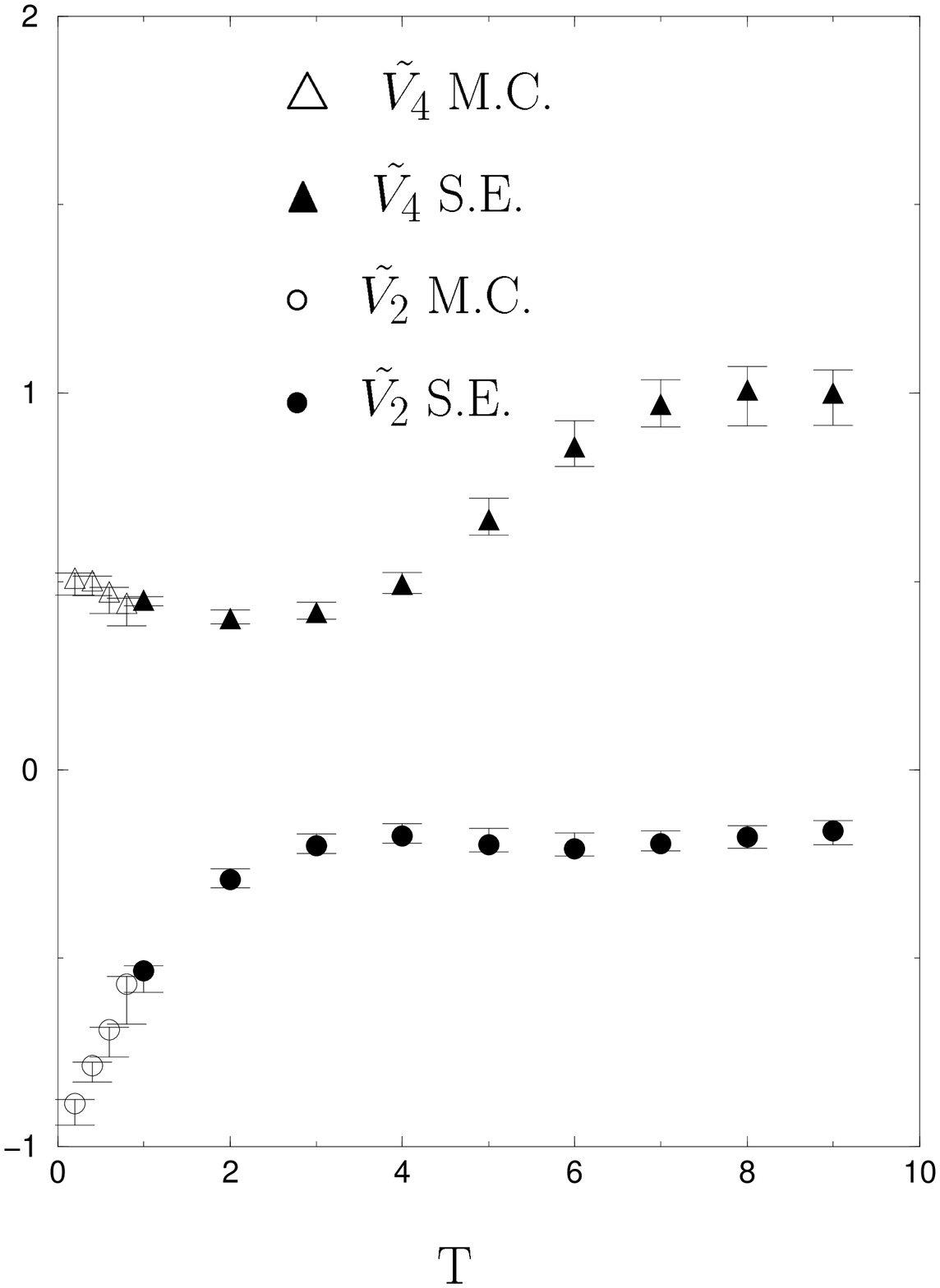,height=9cm,width=.9\linewidth,angle=0}
\end{center}
\noindent 
Fig.[1b] Same as Fig.[1a], for $\tilde{v_2}$ and $\tilde{v_4}$.
\end{figure}
on the quantum potential $\tilde{V}$ but also on the quantum mass $\tilde{m}$.
We define the quantum instanton as the classical instanton of the quantum action $\tilde{S}$. It may happen that the classical potential has a double well shape and the quantum potential has not. Then a classical instanton would exist, but no quantum instanton. For the previous double-well potential the classical instanton and the quantum instantons for different imaginary times $T$ are shown in Fig.[2].
Making a distinction between the quantum potential with
(a) a single minimum versus the case of having (b) degenerate double minima
has a well known analogue in quantum field theory, e.g. in one- or multi-component $\phi^{4}$ field theory. In such field theory a degenerate vacuum may or may not occur, depending on the parameters of the interacting field theory. If a degenarate vacuum exists, the vacuum expectation value 
\begin{figure}
\begin{center}
\epsfig{figure=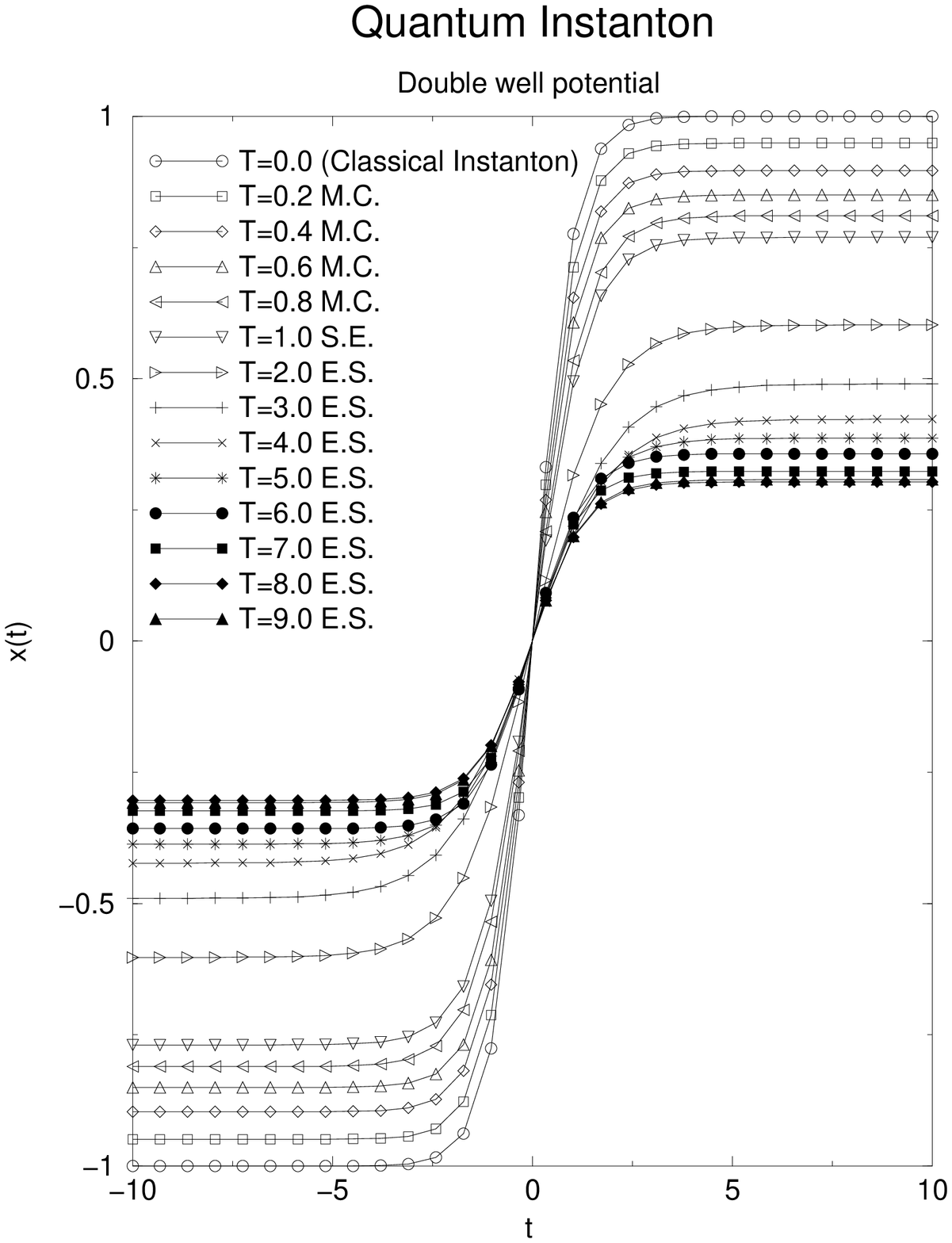,height=9cm,width=.9\linewidth,angle=0}
\end{center}
\noindent
Fig.[2]. Classical instanton ($T=0$) and quantum instantons 
for different imaginary times $T$ (inverse temperature $\beta$).
\end{figure}
$\langle 0 | \phi | 0 \rangle$ takes a value such that the field "sits" 
at a minimum of the potential of the interacting field theory (broken phase, spontaneous symmetry breaking).
It may happen that the noninteracting field theory has a degenerate vacuum, but the interacting field theory does not. Then the field is said to be in a symmetric (unbroken) phase. 
A careful numerical study of the phase structure of this model using a space-time lattice has been performed by L\"{u}scher and Weisz \cite{Luscher}. Because the interacting field theory in Q.M. corresponds to the quantum action, we see that in Q.M. the quantum action is the object to be analyzed for these matters.

\section{Quantum chaos}
The problem with quantum chaos is similar to the problem of quantum instantons. Thus we have suggested in Refs.\cite{Jirari:b,Jirari:c} 
to characterize quantum chaos via the quantum action. 
Let us reflect what this means.
In classical chaos, one is used to construct trajectories in phase space
and analyze non-linear dynamical behavior from such information.
Geometrical concepts like Lyapunov exponents, Poincar\'{e} section etc. are built on this.
Now consider the quantum action. Suppose we have constructed it for some system, which is known to be classically chaotic.
Then we can solve the Euler-Lagrange equations of motion for the quantum action. From that we can construct its phase space. We can follow 
trajectories starting from two initial conditions close in phase space.
This is mathematically well defined. But what does it mean in physics terms?
That brings us to the question: Is there some reality to the quantum action or is it only a mathematical picture?
Is there a particle, the dynamics of which is described by the trajectories of the quantum action?

To all those questions we can not offer a definite answer. As a tentative to an answer let us offer the following comments:
In physics the concept of an effective theory and also the concept of quasi particles is well known. Above we have exemplified an effective theory as a model which describes the important degrees of freedom at some low energy scale. Maybe more familiar is the concept of renormalisation describing the propagation of a particle in a solid versus the same particle in empty space.
The presence of the solid medium modifies the properties of the propagating particle, e.g. inertial mass, effective charge due to phonon effects, etc.
Quasi-particles are well known in condensed matter physics, e.g. excitons, polarons, Cooper pairs etc.
In this sense, we suggest to interpret the quantum action as an effective theory, which describes the dynamics of an effective or quasi-particle.

That brings us to the next question: Is this quasi-particle real? Or better: Is it observable? We are familiar in quantum physics with the wave function, which is not an observable.
However, in interference experiments one can measure the phase difference in the wave function of a particle when it travels different pathes. Or to give a more familiar example:
Scattering phase shifts in a scattering experiment are determined from experimentally measured scattering cross sections. Those phase shifts reflect the change of phase of the wave function of a scattered particle having undergone an interaction with the target.
In scattering experiments, e.g. in nucleon-nucleon scattering in the energy range of a few $MeV$, one measures scattering phase shifts as a function of energy and quantum numbers like angular momentum. They serve to determine the unknown nucleon-nucleon interaction and parametrize it by some low energy effective potential (Bonn potential, Paris potential, etc.). 
In the same sense we would like to suggest that the parameters of the quantum action can be measured experimentally. We propose that this can be done
in an interference or a scattering experiment.
Scheme:
\begin{eqnarray}
&&\mbox{nn scattering experiment} \Longleftrightarrow
\mbox{differential cross section} \Longleftrightarrow
\mbox{phase shifts} \Longleftrightarrow
\nonumber \\
&&\mbox{nn potential} \Longleftrightarrow
\mbox{classical action} \Longleftrightarrow
\mbox{quantum action} ~ .
\end{eqnarray}
Because the quantum action parametrizes the Q.M. transition amplitude and 
the transition amplitude in the limit $T \to \infty$ and suitable boundary conditions (plane waves) determines the $S$-matrix in a scattering process, 
we suggest that it should be possible to extract the quantum action directly from a scattering experiment.
\begin{eqnarray}
&&\mbox{nn scattering experiment} \Longleftrightarrow
\mbox{quantum action} ~ .
\end{eqnarray}
\begin{figure}
\begin{center}
\epsfig{figure=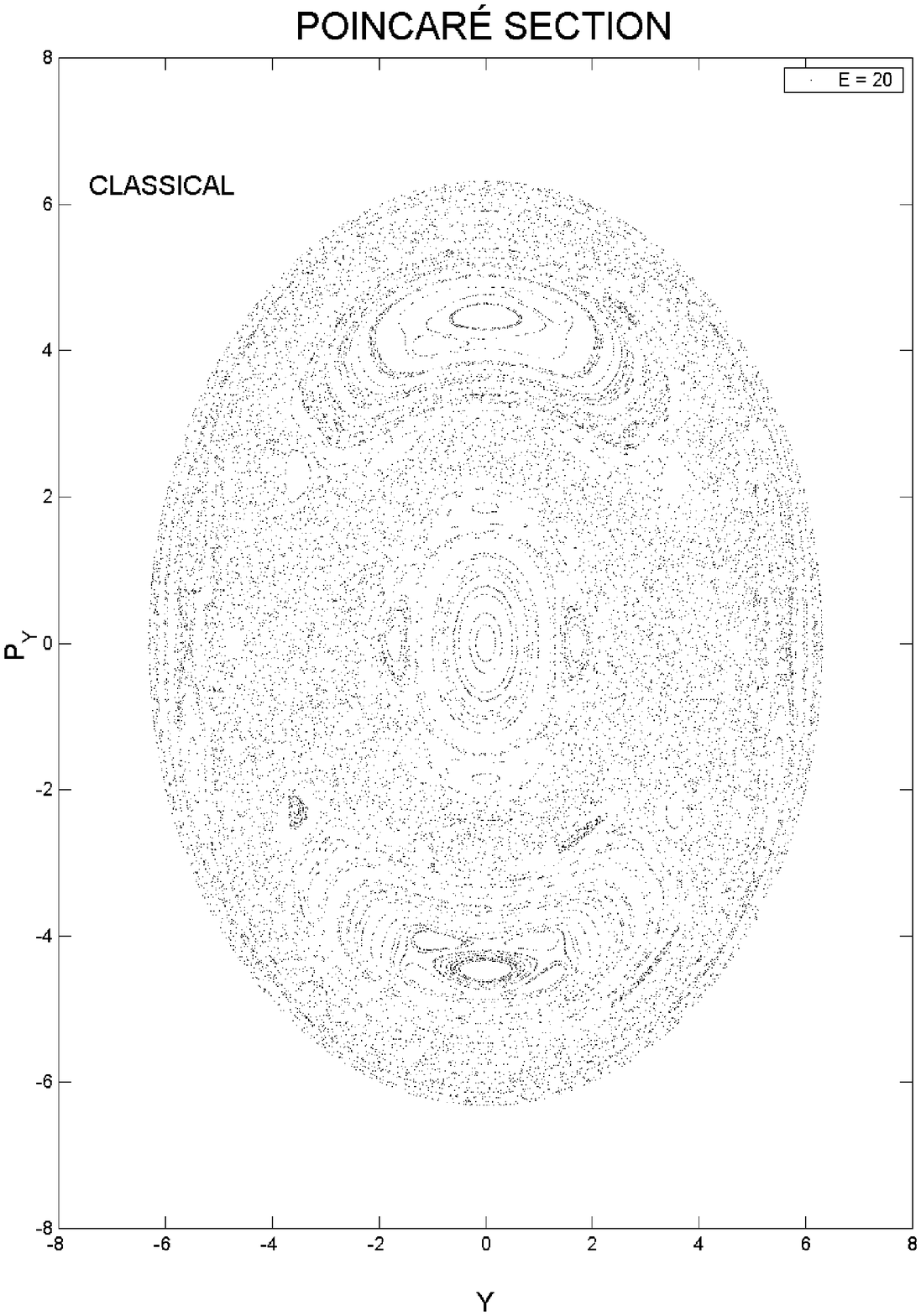,height=9cm,width=.9\linewidth,angle=0}
\end{center}
\noindent
Fig.[3a]. 2-D anharmonic oscillator. Poincar\'e sections of classical action. 
Energy $E=20$. 
\end{figure}
\begin{figure}
\begin{center}
\epsfig{figure=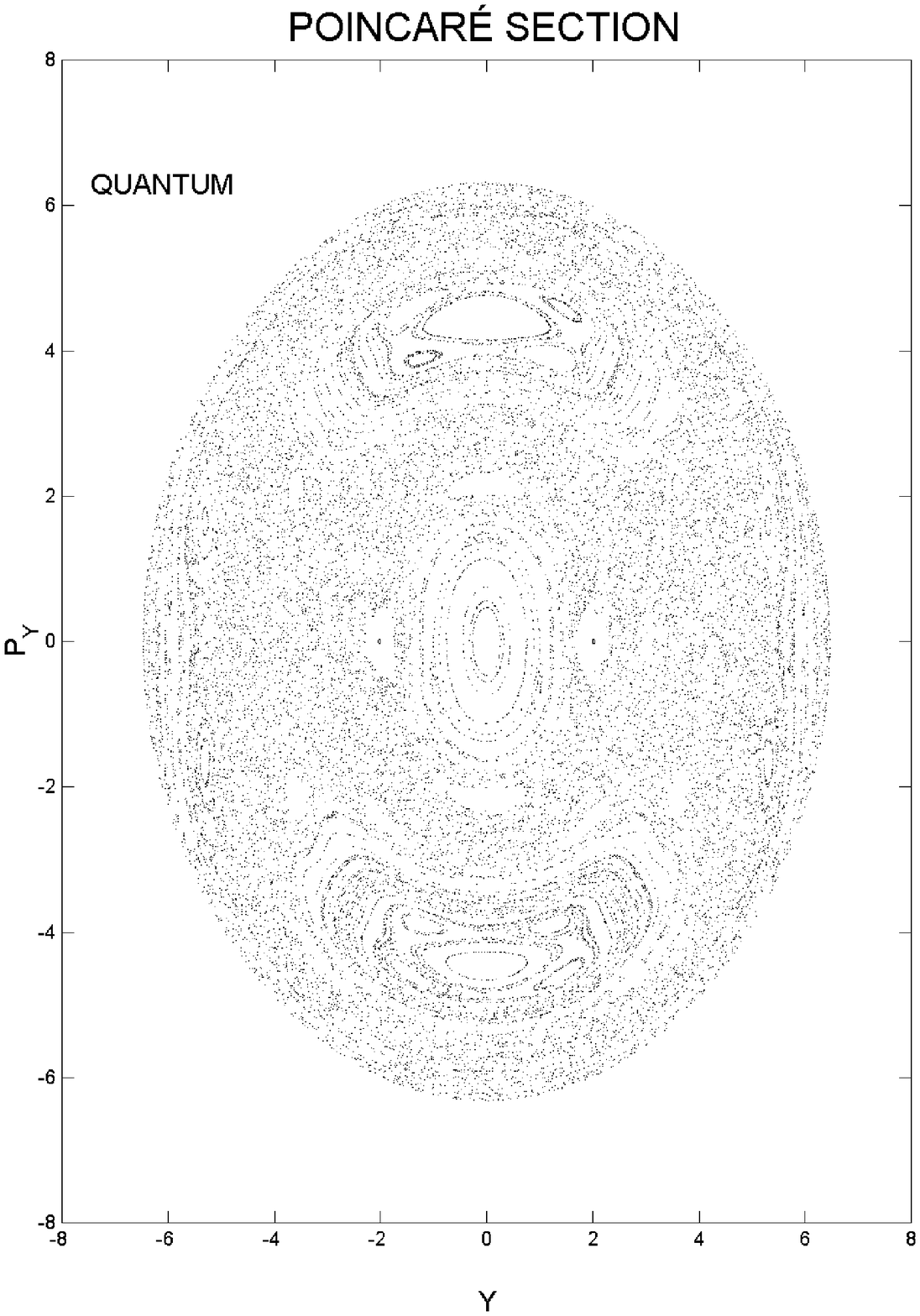,height=9cm,width=.9\linewidth,angle=0}
\end{center}
\noindent
Fig.[3b]. Same as Fig.[3a], Poincar\'{e} section of quantum action at temperature $\tau=0.25$. $E=20$.
\end{figure}
Now let us consider a numerical study of a classical conservative (non-dissipative) chaotic system, and its Q.M. counterpart.
Because in 1-dim a conservative systems with time-independent Hamiltonian is integrable, it does not generate classical chaos.
Hence it is not appropriate to search for quantum chaos in such system.
However, there is a number of 2-dim Hamiltonian systems,
which are known to display classical chaos, e.g. the anharmonic oscillator, the K-system, the Henon Heiles system, the Paul-trap, etc. 
Pullen and Edmonds \cite{Pullen} have suggested that the anharmonic oscillator 
is a model exhibiting classical chaos which is convenient from the numerical. It is defined  by the following classical action, 
\begin{eqnarray}
S &=& \int_{0}^{T} dt ~ \frac{1}{2} m (\dot{x}^{2} + \dot{y}^{2}) 
+ V(x,y), ~~~ V(x,y) = v_{2}(x^{2} + y^{2}) + v_{22} x^{2}y^{2} 
\nonumber \\
m &=& 1 
\nonumber \\
v_{2} &=& 0.5
\nonumber \\
v_{22} &=& 0.05 ~ .
\end{eqnarray}
We work in imaginary time and use the convention $\hbar=k_{B}=1$.
For the corresponding quantum action we made the following ansatz,
compatible with time-reversal symmetry, parity conservation and symmetry under exchange $x \leftrightarrow y$,
\begin{eqnarray}
\tilde{S} &=& \int_{0}^{T} dt ~ 
\frac{1}{2} \tilde{m} (\dot{x}^{2} + \dot{y}^{2}) 
+ \tilde{V}(x,y), ~~~
\nonumber \\
\tilde{V} &=& \tilde{v}_{0} 
+ \tilde{v}_{2} (x^{2} + y^{2}) 
+ \tilde{v}_{22} x^{2}y^{2} 
+ \tilde{v}_{4} (x^{4} + y^{4}) ~ .
\end{eqnarray}
We have determined numerically the parameters of the quantum action 
for transition times from $T=0$ up to $T=4$ (corresponding to temperatures 
$\tau = \infty$ and $\tau=1/4$, respectively). At $T=4$ a regime of asymptotic stability has been reached. 
We have computed Poincar\'e sections from the classical action 
and from the quantum action for a variety of temperatures. 
The equations of motion of the quantum action have been solved using a 4-th order Runge-Kutta algorithm, and Henon's algoritm was used to compute the Poincar\'e sections. 
We computed the Poincar\'{e} sections at different energies from the quantum actions at different temperatures.  
As an example we compare in Fig.[3] for energy $E=20$ classical Poincar\'{e}
section from the classical action with the Poincar\'{e} section from the quantum action at temperature $\tau=0.25$ (corresponding to $T=4$).
Because the quantum action goes over to the classical action in the limit 
$T \to 0$, one can interpret the classical system 
as the quantum system in the limit of temperature $\tau \to \infty$.
Thus the above comparison can be viewed as comparing quantum Poincar\'{e} sections at temperature $\tau = \infty$ and $\tau = 1/4$.
In general one observes that the quantum system also displays chaos, and the Poincar\'{e} sections are slightly different from those of the classical action. Like in the classical case also in the quantum system the "amount" of chaoticity increases with increasing energy.   
Moreover, the difference between classical and quantum Poincar\'e sections becomes more accentuated with increase of energy.

\vspace{1.0cm}

\noindent {\bf Acknowledgements} \\ 
H.K. and K.M. are grateful for support by NSERC Canada. 
X.Q.L. has been supported by NSF for Distinguished Young Scientists of China, by Guangdong Provincial NSF and by the Ministry of Education of China.

\end{document}